# A Health Monitoring System Based on Flexible Triboelectric Sensors for Intelligence Medical Internet of Things and its Applications in Virtual Reality

*Junqi Mao, Puen Zhou, Xiaoyao Wang, Hongbo Yao, Liuyang Liang,Yiqiao Zhao, Jiawei Zhang *,Dayan Ban* and Haiwu Zheng**

J. Q. Mao, P. E. Zhou, X. Y. Wang, H. B. Yao, L.Y. Liang, Y.Q. Zhao, J. W. Zhang, Prof. H. W. Zheng

Henan Province Engineering Research Center of Smart Micro-nano Sensing Technology and Application, School of Physics and Electronics, Henan University, Kaifeng 475004, P. R. China

E-mail address: Jiawei Zhang: 10110130@vip.henu.edu.cn

E-mail address: Haiwu Zheng: zhenghaiw@ustc.edu

Prof. D. Y. Ban

Waterloo Institute for Nanotechnology and Department of Electrical and Computer Engineering, University of Waterloo, Waterloo N2L3G1, ON, Canada

E-mail address: Dayan Ban: dban@uwaterloo.ca

Abstract: The Internet of Medical Things (IoMT) is a platform that combines Internet of Things (IoT) technology with medical applications, enabling the realization of precision medicine, intelligent healthcare, and telemedicine in the era of digitalization and intelligence. However, the IoMT faces various challenges, including sustainable power supply, human adaptability of sensors and the intelligence of sensors. In this study, we designed a robust and intelligent IoMT system through the synergistic integration of flexible wearable triboelectric sensors and deep learning-assisted data analytics. We embedded four triboelectric sensors into a wristband to detect and analyze limb movements in patients suffering from Parkinson's Disease (PD). By further integrating deep learning-assisted data analytics, we actualized an intelligent healthcare monitoring system for the surveillance and interaction of PD patients, which includes location/trajectory tracking, heart monitoring and identity recognition. This innovative approach enabled us to accurately capture and scrutinize the subtle movements and fine motor of PD patients, thus providing insightful feedback and comprehensive assessment of the patients' conditions. This monitoring system is cost-effective, easily fabricated, highly sensitive, and intelligent, consequently underscores the immense potential of human body sensing technology in a Health 4.0 society.

# 1. Introduction

The demand for intelligent healthcare systems and services for the elderly is increasing with the aging global population. According to the World Health Organization, the proportion of the world's population aged 60 years or over will nearly double from 12% to 22% from 2015 to 2050 [1]. Parkinson's disease (PD) is a chronic neurodegenerative disorder that affects the elderly, with a prevalence of 1,663 per 100,000 in those aged 80 and older. Among them, 4% of PD patients are unable to live independently without caregiver assistance [2]. Continued development of wearable sensor technologies has the potential to provide clinicians with holistic and objective information about the patient's health status, facilitating personalized interventions, alleviate symptoms, and slow the disease progression [3, 4]. In this system, wearable sensors are attached to the skin or worn on the body to collect various physiological and motion information, enabling the identification of different human postures. For instance, Chen et al. introduced a body area sensor network consisting of edible triboelectric hydrogel sensors for monitoring omnidirectional baby motion [5]. Lau et al. developed a self-powered strain sensor based on graphene oxide-polyacrylamide hydrogel to monitor subtle human motions, including gait motions [6]. By integrating these wearable sensor systems with AI algorithms, it becomes possible to digitize human activities and generate data-driven insights for health status and activity management. By these attempts, the burden on caregivers and healthcare professionals can be significantly alleviated.

The IoMT integrating IoT technology and medical applications can enable the realization of precision medicine, intelligent healthcare, and telemedicine in the digital and intelligent age. Common types of wearable sensors used in IoMT mainly include photovoltaic sensors, piezoelectric sensors and resistive sensors, capacitive sensors.

The large number of sensor nodes in IoMT make regular battery replacements costly and inconvenient in some scenarios. Additionally, the massive amount of data generated by sensors in IoMT requires efficient processing and computation. Consequently, there is an urgent demand for the development of intelligent IoMT systems with sustainable power sources. The Metaverse, a network of interconnected virtual reality spaces with which users can interact and experience through wearable devices, offering a novel opportunity to enhance the quality of life for PD patients [7, 8]. Wearable sensors in the IoMT systems can be used to project human movements, emotions, physiological signals and environmental interactions from the real world into the virtual world, further transforming user experiences in human-machine interaction, intelligent healthcare, immersive virtual reality (VR) and augmented reality [8, 9]. Through the Metaverse, PD patients can transcend their physical limitations and engage in more free interactions and experience [10, 11]. Although the metaverse offers new possibilities for the future health monitoring, it faces practical challenges. One of these challenges is the limitation posed by the cost, maintenance, and power supply of wearable sensor nodes.

A potential solution to address the aforementioned challenges lies in the triboelectric sensors, from which the electrical signals produced by triboelectric nanogenerators (TENG) can be employed to sense human motion and physiological activities [12, 13]. Wearable triboelectric sensors can capture rich physiological information such as blood pressure, temperature, vital signs and so on, which provide a comprehensive understanding of the wearer's health status to support clinical decision making while reducing the workload of human caregivers [14]. In addition, advanced deep learning model favors to achieve more immersive interaction between PD patients and the Metaverse, which requires the collection of massive volumes of data from PD patients. Namely, the model processes multimodal information from wearable sensors and creates a final information model to project all of the information of a human in the real world into their digital avatar in the Metaverse. [15-17]

In this study, we propose the synergistic fusion of flexible triboelectric sensors and deep learning-assisted data analysis to yield a robust and intelligent IoMT system [18-20]. This system enhances the interaction experiences of patients within the Metaverse environment [21-23]. The designed flexible triboelectric sensors (FTES) with wearable characteristics are composed of nylon, flexible Polyethylene terephthalate (PET)/Indium Tin Oxide (ITO) and polyimide (PI), demonstrating good biocompatibility and high sensitivity. Four FTES are attached to commercial wristbands to enable accurate human posture recognition. These sensors subjected to relative motions generate distinct triboelectric signals depending on limb movements. In parallel, we devised a hybrid neural network that incorporates an attention mechanism, amalgamating Convolutional Neural Networks (CNN) and Bidirectional Long Short-Term Memory (BiLSTM) for posture recognition. The CNN-BiLSTM-Attention model relies exclusively on four sensors to accurately identify eight distinct human postures, achieving an impressive average accuracy of 97.3%. Moreover, by the aid of deep learning-assisted data analysis, we can determine user identities by classifying posture-induced outputs within a pre-trained deep learning model. By leveraging this innovative design and integrating the model with artificial intelligence (AI), we developed an intelligent healthcare monitoring system, capable of concurrent posture recognition, identity verification, and health monitoring functions. The versatility of this system makes it suitable for a wide range of intelligent healthcare monitoring and VR interaction applications.

## 2. Results and Discussion

This intelligent IoMT monitoring system, as illustrated in Figure 1a, is a flexible solution for comprehensive monitoring of patients with PD. It can be seamlessly deployed in various settings, from community environments to hospital facilities. It facilitates a broad spectrum of medical monitoring and VR interactions, and thereby enhances the quality of healthcare delivery. The system integrates time-domain analysis and deep learning-assisted analysis to perform multiple functions concurrently, such as heart rate monitoring, posture recognition, VR interaction, real-time positioning, and identity recognition. The system employs flexible triboelectric sensors, which were positioned on the limbs to capture comprehensive body movements. These sensors monitor the overall movement patterns and behaviors of patients, providing invaluable data for clinical research and caregiving.

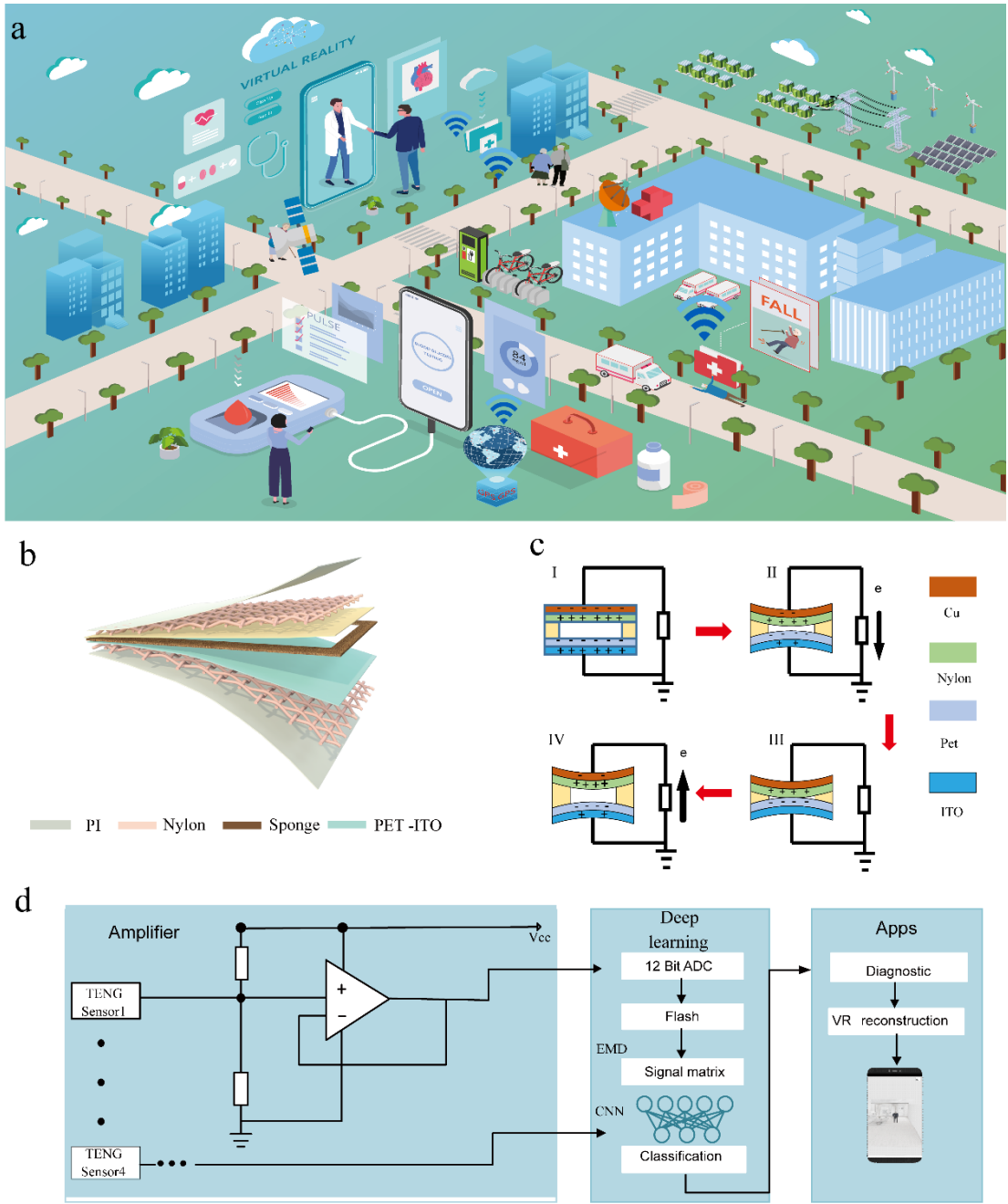

**Figure 1.** Overview of the intelligent IoMT monitoring system and detailed sensor structure design. a) The schematic diagram of the intelligent IoMT monitoring system with both time-domain data processing and DL-assisted data analytics to achieve real-time position sensing and VR applications. b) A diagrammatic sketch of a wearable and flexible TENG-based sensor. c) working mechanism of the TENG sensor. d) A schematic diagram of deep learning assisted body area triboelectric flexible sensor for real-time posture monitoring system.

The working principle of the triboelectric sensor to generate electrical signals is illustrated in Figure 1c. The relative motion between the nylon and PET, induced by limb movements, facilitates the contact-separation process. In *stage* I, when the body is at rest, the charges remain unchanged, with the nylon and PET films inducing corresponding positive and negative charges on the copper electrodes, respectively. In *stage* II, the nylon can be propelled by the arm muscle to approach the PET. In this case, the induced charges on the copper electrodes decrease, and current flows from the ITO electrode on the PET side to the Cu electrode on the nylon side. When the nylon and PET are in full contact at *stage* III, the current decreases to zero, and the charges induced on the electrodes reduces to nearly zero. In *stage* IV, as the nylon moves away from the PET, the charges induced on the Cu electrode increase, and the current flows from the Cu electrode to the ITO electrode, completing an entire cycle. The COMSOL simulation demonstrates the distribution of the electric field in the sensor under different pressure release states, as shown in Figure S1of the *Supporting Information*. Figure 1d shows an adaptive motion-monitoring system, from which the body sensor network can be readily integrated a microcontroller unit, deep-learning algorithms, and mobile VR display terminals. This system can achieve high-precision posture recognition and interaction with different postures of PDs in real time.

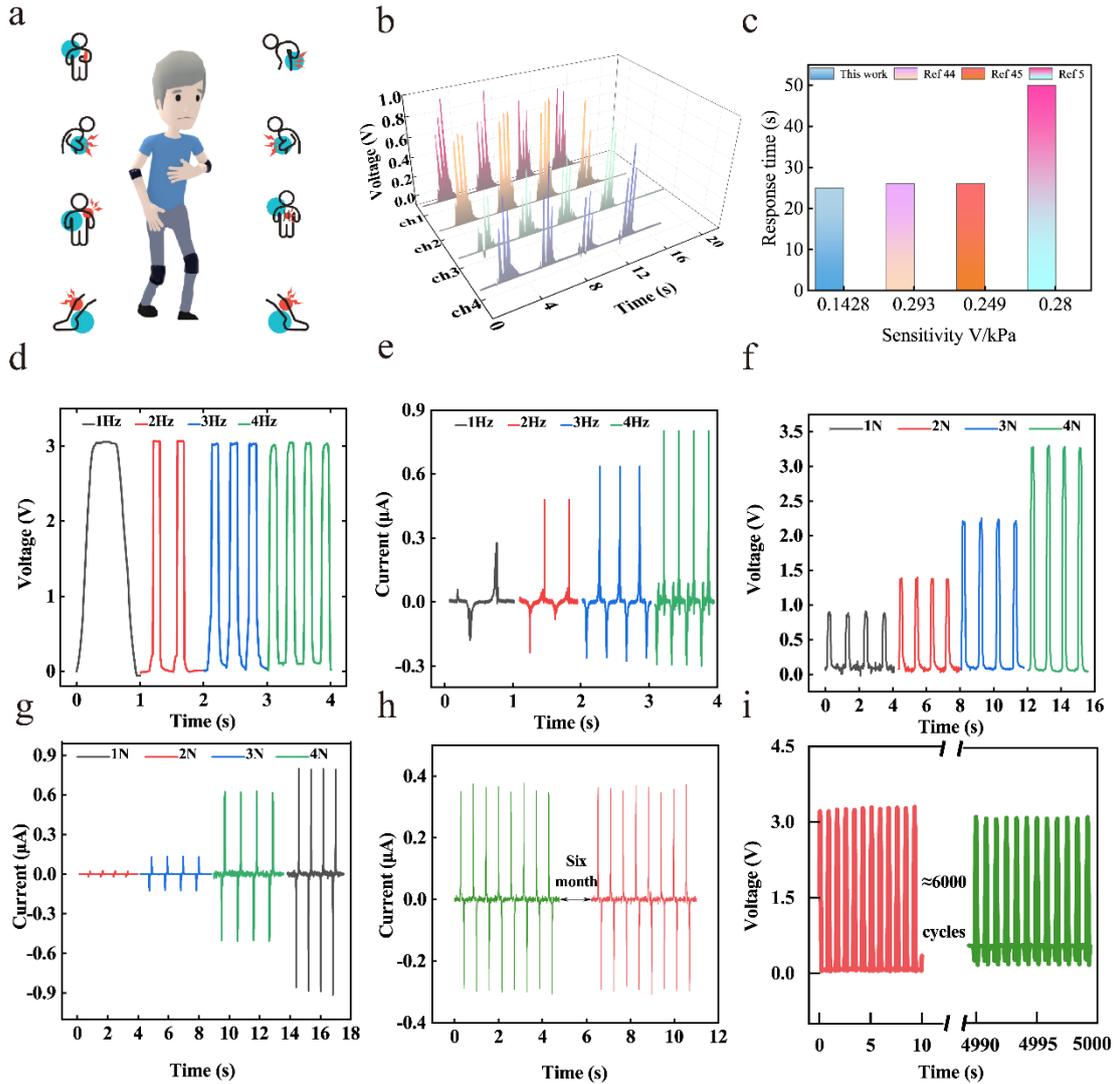

**Figure 2.** Characterization of the FTES. a) Schematic of integrating FTES on wrist band and kneecap. b) Schematic of the user stepping and the corresponding output curve of the four-channel FTES. c) Characteristic comparisons between previously wearable sensors and the current sensor. d-e) The open-circuit voltage and short-circuit current output of the sensor at the pressing frequency (from 1 Hz to 4 Hz) with a force of 4N. f-g) The open-circuit voltage and short-circuit current obtained by pressing the sensor at 1 Hz frequency with different forces (from 1 N to 4 N). h) Long-term stability of the triboelectric sensor after storage in the dry environment for six months. i) Mechanical durability test that lasted for ≈6000 cycles. Inset: The $V_{OC}$ output signals generated for the beginning (0–10 s), in the middle part (4990–5000s).

PD is a neurodegenerative disorder that affects the patient's motor system, causing symptoms such as bradykinesia, tremor, rigidity, and gait balance problems. Thus, it is necessary to monitor the movement patterns of patients with this disease by the FTES, which could sense the motion signals of the joints and provide feedback for diagnosis and treatment. The FTES was attached to the elbow and knee joints, which were integrated with safety band for support and comfort, as shown in Figure 2a. Figure 2b shows the voltage signals collected by the four sensors during a stepping activity. To highlight differences in signal waveforms and facilitate the signal processing, the amplitude interference is eliminated by normalization. This FTES demonstrates a high performance with a sensitivity of 0.1428 V/kPa and a response time of 26 ms, which are calculated from Figures S4 and S5 of Supporting Information, respectively. From Figure 2c, it can be seen that the fabricated FTES has many overall performances compared with previously reported triboelectric sensor in the aspect of response time and sensitivity. In addition, as seen in Table 1, the stability, identification accuracy and response time of the fabricated FTES are comparable or superior to the capacitive and piezoelectric sensors. The open-circuit voltage ($V_{OC}$) and short-circuit current ($I_{SC}$) of the triboelectric sensor are vital for posture recognition. The sensor fixed onto a linear motor was pushed periodically at various frequencies and forces. By increasing the application frequency from 1Hz to 4Hz under constant pressure of 4N, the $V_{OC}$ and $I_{SC}$ of the sensor were tested and shown in Figures 2d and 2e. The $I_{SC}$ increased from 0.3 μA to 0.8 μA with the increasing frequency, while the voltage remained nearly constant. As the applied force increased from 1 N to 4 N under a constant frequency of 1 Hz, the $V_{OC}$ increased from 0.9 V to 3.5 V and the $I_{SC}$ increased from 0.15 μA to 0.8 μA with the increase in force, as indicated in Figures 2f and 2g. The durability and stability of signal for the FTES were evaluated by measuring its output voltage after six months and 6000 cycles under compressive stress of 2 N at 1 Hz. It seems that the $V_{OC}$ of the FTES was stable during the cyclic loading, as illustrated in Figure 2h.

Furthermore, the signal-to-noise ratio (SNR) of the FTES was calculated. The SNR can be defined as

$$\text{SNR} = 10\log_{10}\frac{P_s}{P_n}$$

where $P_s$ and $P_n$ are the average power of collected motion signal and noise signal, respectively. The result reflects that the FTES has an SNR of 31.2 dB, which is higher than that reported in the previous literature [5]. A more in-depth derivation can be found in Note S1 of Supporting Information. To test the flexibility of the sensor, different bending angles of the FTES was performed by attaching it to the finger, as shown in Figure S6, from which it seems that the FTES can generate significant electrical signal even though at bending angle up to 5°. Consequently, the FTES possesses a collection of remarkable attributes such as high sensitivity, swift response time and durability. These qualities make it suitable for accurate, long-term health monitoring of PD patients, thereby enhancing the effectiveness of the care provided.

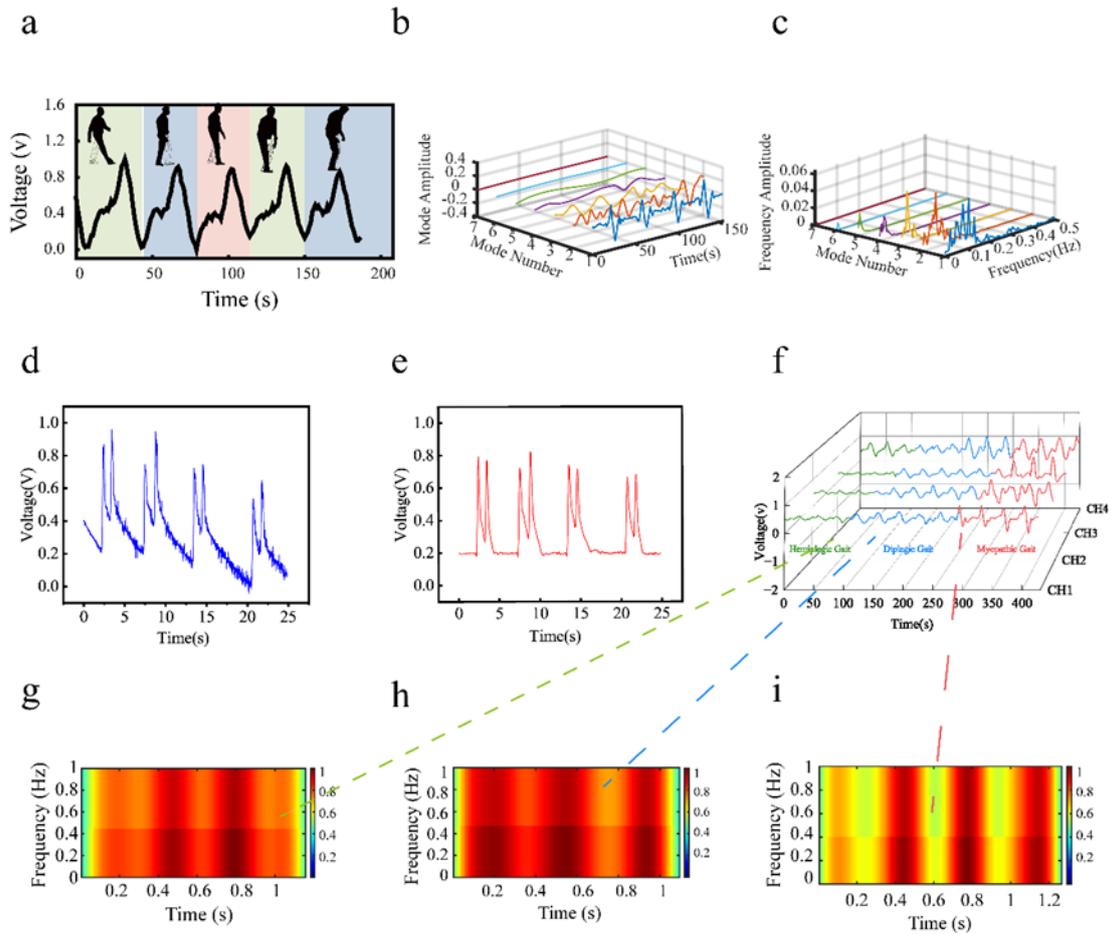

**Figure 3**. An EMD Denoising and Feature Extraction Algorithm. a) The typical 5 phases in a normal walking process. b-c) EMD decomposition of walking signal in the time domain and the frequency domain. d) Normalized motion signals generated by body motion. e) Stable and smooth motion signals after wavelet reconstruction. f) Real-time monitoring of the normalized voltage signals of Parkinson's disease, including hemiplegic gait, diplegic gait and myopathic gait. g-i) STFT analysis of the leg motion signals of 'Hemiplegic Gait', 'Diplegic Gait' and 'Myopathic Gait'.

The FTES could monitor the movement patterns and posture behaviors of patients, providing invaluable data for clinical research and caregiving. Figure 3a presents the posture consisting of five distinct phases when an imitated PD patient walks and the corresponding output electrical signal captured by the FTES. The electrical signals derived from the trembling of five fingers of an imitated PD patient were tested, as shown in Figure S7. The above-mentioned output signals could be adopted for deep learning-assisted identity recognition, which is accomplished through the extraction and classification of subtle features in the sensor signals. Because of uncontrollable factors such as environmental interference, the FTES may suffer from instability and baseline drift during long-term operation, causing distorted data and obstructs classification and posture recognition. Therefore, intelligent signal processing with the function of denoising and baseline calibrating is essential for accurately identifying the postures of PD patient.

We utilized the Empirical Mode Decomposition (EMD) algorithm, a powerful tool for analyzing various nonlinear and non-stationary signals without the requirement to preset basis functions [24, 25]. The EMD algorithm can capture the local features and temporal variations of a signal with larger accuracy and intuitiveness, which can decompose the signal into intrinsic mode functions (IMFs) components [26, 27]. The fundamental principle of the EMD algorithm is to construct upper and lower envelopes using local extreme value of the signal, calculate the mean values of these envelope curves, subtract the mean values from the original signal, and obtain the first IMF. [28, 29]. This process is repeated until a function that satisfies the IMF definition is secured. The first IMF is then subtracted from the original signal to obtain a residual signal. The same process was performed on the residual signal to derive the second IMF. This process continues until the residual signal is a monotonic function or falls below a preset threshold. Ultimately, the original signal can be expressed as the sum of several IMFs and a residual signal [30, 31]. A detailed decomposition process of the EMD algorithm can be found in Note S2 of the Supporting Information.

Figures 3b and 3c illustrate the different modal components of the signal decomposition from Figure 3a and their expression in the time and frequency domain, respectively. Each IMF exhibits a locally varying amplitude and frequency, reflecting the fundamental characteristics of the signal. As demonstrated in Figure 3d and Figure 3e, the reconstructed signal retains the signal edge while eliminating signal oscillation and baseline drift. The detailed derivation process can be found in Note S3 of the Supporting Information. Figure 3f presents the normalized voltage of an imitated PD patient performing three different types of activities — 'Hemiplegic Gait', 'Diplegic Gait', and 'Myopathic Gait' while wearing the FTES onto limbs, showcasing activity-monitoring capabilities of the FTES. Furthermore, taking 'Hemiplegic Gait' as an example, the signals derived from two distinct human arm movements were analyzed and expressed in the frequency domain (Figure S9a). The frequency-domain curves originated from the channel 1 are shown in Figure S9b, where the three activities are easily distinguishable by the overall amplitude and frequency of the output signals. However, the properties of the variations cannot be fully depicted by merely employing information in the individual frequency or time domain. Therefore, short-time Fourier transform (STFT), an advanced mathematical method derived from the discrete Fourier transform (DFT) that can explore the instantaneous frequency and the instantaneous amplitude of local waves with time-dependent features was adopted. Figures 3g, 3h, and 3i show the STFT transforms of the three different gait patterns, highlighting the variations in frequency characteristics among them and thus paving the way for subsequent deep learning.

The rapid advancement of deep-learning technology is gradually changing the landscape of intelligent healthcare. This technology, capable of learning rules from massive data and processing vast amounts of data at a speed and accuracy beyond human capability, can offer proactive insights into unknown details [32]. However, traditional machine learning and CNN models can only be employed to extract spatial features of signals and easily ignore temporal features, causing the decrease of recognition accuracy. [33]. To overcome this limitation, we introduce a hybrid CNN-BiLSTM-Attention model, representing a novel deep-learning framework for human activity recognition [34-37]. The model effectively extracts both spatial and temporal features of input signals, learning the long-term dependencies and importance weights of hidden states to provide high accuracy and robustness in human activity recognition [38, 39]. The CNN layer plays a crucial role in denoising, dimensionality reduction and enhancing the representation ability of the model. For the CNN backbone, we fine-tuned the number of filters, kernel size, and convolutional layers to optimize recognition performance. As demonstrated in Figures 4a, 4b, and 4c, the CNN model with 64 filters, a kernel size of 32, and 2 convolutional layers delivers the best overall accuracy. Figure 4d presents a schematic diagram of the feature extractor structure within the model. BiLSTM, an extension of LSTM (Long Short-Term Memory), includes an additional backward LSTM layer. This layer processes the reversed version of the input sequence. BiLSTM can leverage both the past and the future information for each point in the input sequence by training a forward LSTM layer and a backward LSTM layer, respectively.

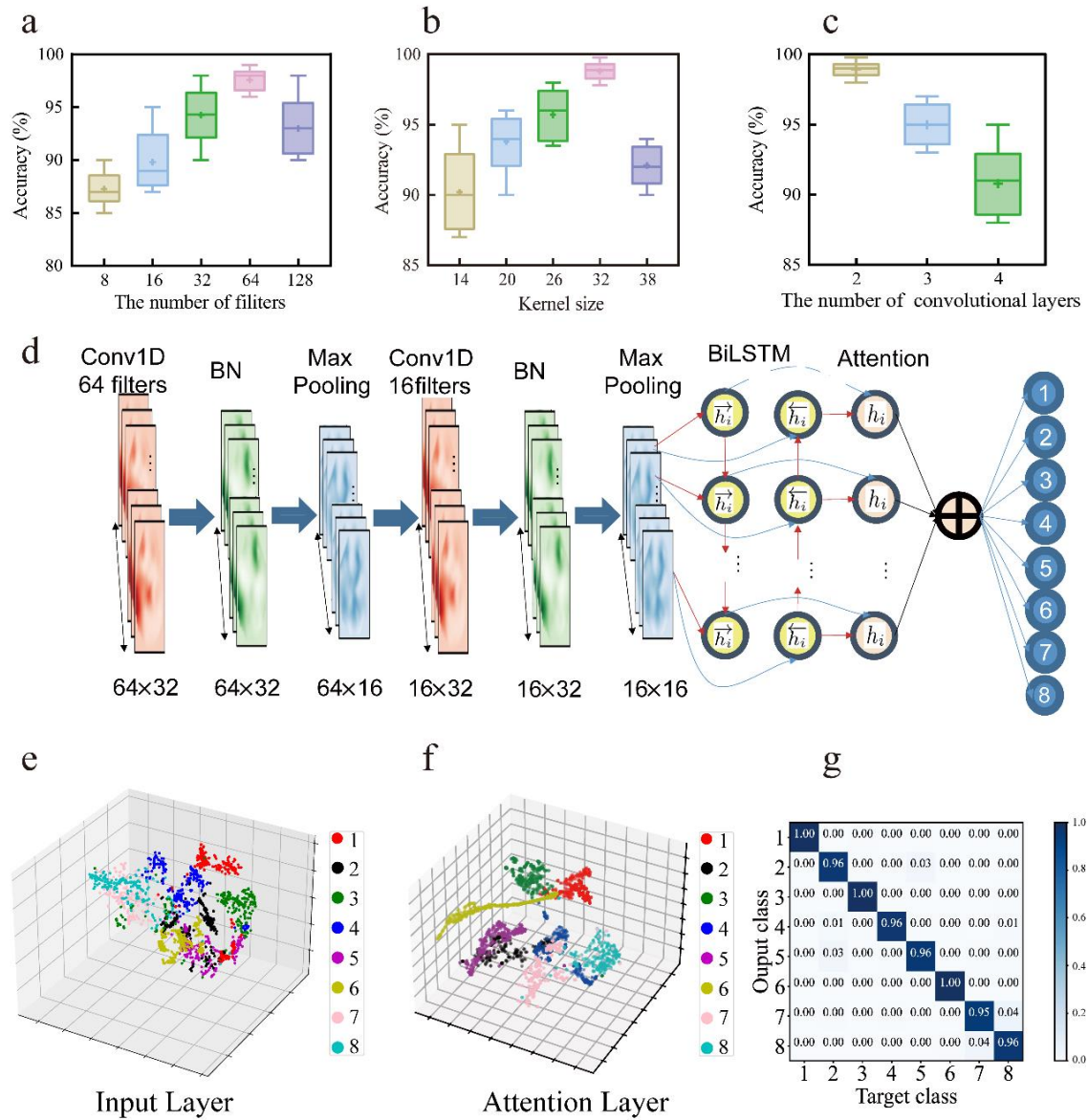

**Figure 4**. PD patient identification based on the smart wrist band, kneecap and deep learning. a-c) The schematic illustration and energy band diagrams under optimization of CNN structure parameters based on accuracy performance by kernel size, filter number, and the number of convolutional layers. d) The detailed structure of the convolutional neural network (CNN) training model. e-f) The distribution of datasets before input layer (e) and after attention layer(f) in a three-dimensional space composed of PC1, PC2 and PC3 axes. The different colored clusters represent the datasets of different postures. (g) Confusion matrix for recognition of eight postures recognition, including Class 1: normal walking; Class 2: jumping; Class 3: falling down; Class 4: hemiplegic gait; Class 5: diplegic Gait; Class 6: running; Class 7: running; and Class 8: Tai Chi.

This unique capability allows BiLSTM to provide better accuracy than traditional LSTM. In the proposed model. Specifically, the model extracts local features through the CNN layer and then feeds them into the BiLSTM layer, enabling the model to learn the temporal dependencies among the features. Consequently, the model effectively captures both the local features and their temporal variations over time. In CNN-BiLSTM-Attention model, the attention mechanism is incorporated following the bidirectional LSTM layer. This allows the model to automatic capture of the most relevant features of the input sequence. It serves as an adaptive feature selection strategy that effectively enhances the model's performance in addressing intricate sequence tasks. Overall, this model substantially enriches the model's comprehension of the input data, empowering it to more effectively capture and leverage critical information, consequently augmenting the overall performance of the model. More details about CNN, BiLSTM and attention could refer to Table 2 and Note S4 in the *Supporting Information*. To illustrate the clustering performance of the CNN-BiLSTM-Attention model structure more effectively, we apply t-distributed stochastic neighbor embedding (t-SNE) in a 3D feature space. Different colors on the profiles denote different categories of postures, with each point projected from high-dimensional data sets to low-dimensional space. Figure 4e and 4f display three t-SNE distributions of eight classes at the input and output layers. Figures S10a and S10b show the distribution of datasets before the CNN layer and after training the BiLSTM layer in a three-dimensional space. The boundaries between the eight classes at the output layer are distinct and exhibit minimal overlap. In the present data set, 532 training samples (80%) and 228 test samples (20%) were used for each action.

As shown in the confusion matrix in Figure 4g, the recognition accuracy of these eight postures reached 97.3%, which has higher accuracy and classification results than those reported in previous literature in term of healthcare of PD patient based on TENG. [6] Figures S8a and S8b illustrate the model training process, highlighting the changes in accuracy and the loss function. Namely, the neural network rapidly attained high recognition accuracy and exhibited excellent stability.

In addition, in the new era of intelligent society, the rapid development of VR/augmented reality technology provides technical support for remote healthcare in the era of Health 4.0, which is a medical care strategy and management model inspired by industry [40]. Under the concept of Health 4.0, more intelligent, automated and personalized healthcare services will be highly anticipated.

Grouped node devices play an important role in Health 4.0, which can be regarded as perception layer in the IoT architecture [41]. Intelligent terminal such as mobile phones and computers, can collect, process and upload patient health data for subsequent analysis and care decisions. For this purpose, a personalized rehabilitation system for PD patients was proposed, whose aims are to provide personalized rehabilitation exercises for PD patients at home by projecting the recognition results from real space into virtual space [42, 43].

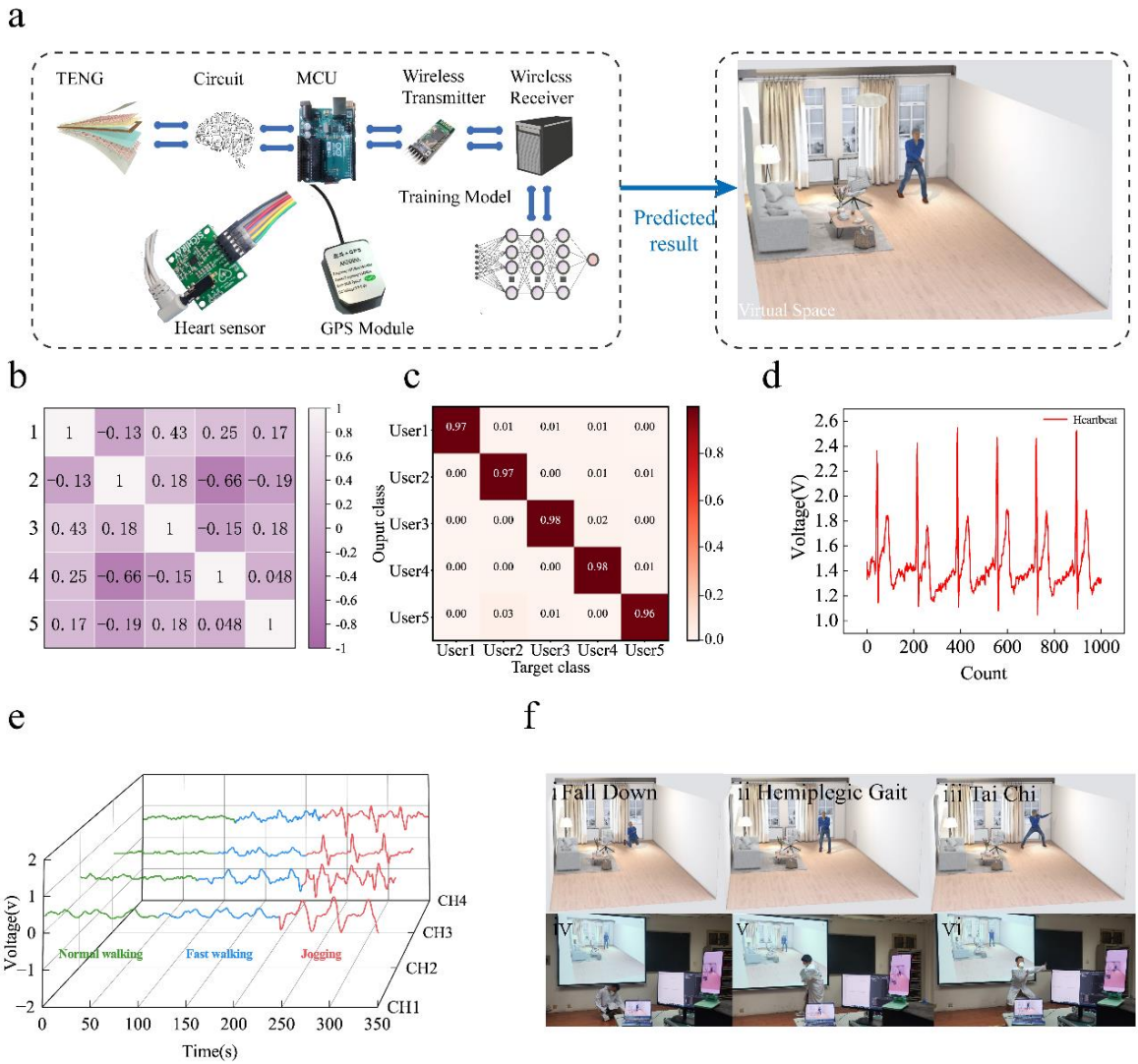

**Figure 5**. Real-time indoor monitoring functions of the FTES enabled by deep learning. a) Schematic of the IoMT. b) Correlation coefficient matrix of five imitative PD patients. c) Confusion matrix for identity recognition for five different imitative PD patients. d) The use of commercial ECG sensors to collect heartbeat signals. e) Normalized voltages of different walking modes, including normal walking, fast walking, and jogging. f) An elderly people (i) falling down, (ii) hemiplegic gait, and (iii) Tai chi in the virtual space with a person (iv) falling down, (v) hemiplegic gait, and (vi) Tai chi in real space.

Figure 5a summarizes the architecture of the IoMT system, which contains the proposed VR application enabled using FTES, signal conditioning circuit, wireless transmitter, GPS module and heart sensor. The IoMT possesses four functions including identity recognition, positioning system, activity detection and heart rate monitoring. To better aid PD patients, especially in outdoor environments, extrasensory information such as location would be highly desirable. To meet this requirement, a commercial global positioning system (GPS) module as one of the functional units was integrated in the IoMT system. An example of GPS data is illustrated in Figure S11. The IoMT system monitors the activity by collecting the PD patient's triboelectric signal and sending it to the deep learning training model via Bluetooth. The triboelectric signals derived from the FTES are collected and processed by Arduino UNO in real time, and these signals can be reflected in VR space. In addition, five imitative PD patients produced highly similar electrical signals, which was revealed by the correlation analysis, as depicted in Figure 5b. It is difficult to accurately recognize different identity by manually extracting shallow features from raw signals. Consequently, we further employed the deep learning approach with exhaustive feature extracting ability. For different patients, 500 signal data points for each channel as the deep learning data were recorded. Subsequently, 50 sets of data samples collected, totaling 2,000 × 50 data points, in which 40 sets of data samples were utilized for model training and 10 sets of data samples were applied for testing. In the confusion matrix shown in Figure 5c, the recognition accuracy of five different patients can reach up to 98%. The IoMT system based on the FTES can achieve high detection accuracies for identity recognition, which demonstrates its great potential for comprehensive monitoring of PD patients apart from the aforementioned posture recognition. We also used a commercial heart rate sensor to monitor the elderly's heart rate and pulse in daily activities and during sleep, as shown in Figure 5d and Video S1, indicating the good extendibility and compatibility of the IoMT system. Figure 5e shows the time-domain signals of PD patients with moderate mobility impairment under three different conditions, including jogging mode, fast walking mode, and normal walking mode. It is evident that the voltages for the above-mentioned three modes are distinct, making them suitable for detecting the motion patterns. The trained posture recognition model was deployed into the server to display the signal waveform and corresponding prediction results of the gaits. At the same time, a corresponding human model will be generated in virtual space, which performs the same action as the predicted posture on multiple terminals. Video S2 in the *Supplemental Materials* provides a demonstration of the elderly real-time motion monitoring capability. In this video, a subject imitates five representative motions including walking, running, hemiplegic gait, Tai Chi, and jumping of PD patient and the FTES was stimulated by the five motions to generate different electrical signals. The photograph originated from the Video S2 is show in Figure 5f, from which real-life images and their corresponding virtual space can be seen. Once a PD patient has fallen down, an abnormal signal generated from the FTES will be projected into the virtual space after deep model processing, facilitating real-time medical monitoring and virtual reality interaction. Unlike traditional camera-based systems that may raise privacy concerns due to video recording, the IoMT system available for indoor monitoring of PD patients has a privacy-preserving attributes. Specifically, the IoMT system can offer an intuitive interface to monitor the PD patient's motion status without invading their privacy by the aid of virtual reality technique. In this system, each terminal can not only operate independently, but also interconnect through a local area network to form a distributed information collection and processing center. The distributed architecture guarantees the safe operation for the IoMT system. Even if one terminal fails or goes offline, the system can continue to remotely monitor the patient's activity status through other terminals. The gesture interaction in VR has a low delay of <1.5 s, which is better than that in the previously reported literature [6]. Furthermore, under the framework of Health 4.0, the IoMT system can achieve in-depth understanding of disease conditions and personalized care through VR technology, which favor to improve the level of intelligence and automation of healthcare services.

By virtue of the IoMT system based on the FTES, the information of health status for PD patients can timely transmit to a centralized data processing and analysis platform. For example, mobile phones may collect patient data such as heart rate and gait pattern, while computers may be used to input and track patient medical records and treatment plans. Through VR technology, medical staff can monitor patient conditions timely in a three-dimensional environment, thus gaining a deeper understanding of patient demands to enable the provision of better, more personalized healthcare.

## 3. Conclusions

Here, we developed an intelligent IoMT system by integrating the FTES and deep learning-assisted data analysis, as a healthcare platform for enabling personalized monitoring and care for the elderly and PD patients. The FTES consisting of a PET-ITO membrane, nylon membrane and spacer was featured by a signal-to-noise ratio of 31.2 dB, a response time of 25 ms and a sensitivity of 0.1428 V/kPa. The IoMT system can achieve various functions including identity recognition with classification accuracy 98% and motion state determination with classification accuracy 97.3% by introducing an advanced attention mechanism-based CNN-BiLSTM hybrid neural network. The GPS and heart-rate sensors in the IoMT system can separately track the user's location and monitor their heart rate in outdoor environments, thereby enabling safe and autonomous all-around monitoring of the elderly without the immediate presence of caregivers. At the same time, the healthcare data collected by the IoMT was projected into VR space and monitored by multiple edge terminals, opening an opportunity for enhancing cyber-human interactions with immersive experience. The IoMT based on the FTES not only provide an important reference for clinicians and healthcare practitioners to develop treatment plans of PD patients, but also promotes the application of triboelectric nanogenerators that capture human biomechanical energy in the realm of artificial intelligence-based medical system.

## 4. Experimental Section

*Preparation of Sensors*: To prepare the FTES, commercial nylon and the PET in commercial PET/ITO were used as friction layers, and copper foil as well as the ITO in the PET/ITO were selected as the electrodes of the FTES. Two squares of $2 \times 2$ cm$^2$ were cut from the purchased PET and nylon. A square of $1 \times 1$ cm$^2$ copper foil was attached to the spacer made from polyurethane sponge. Copper paint wire with a diameter of 70 mm was welded to the copper foil. For yielding a spacer, a square block of 40 mm side length was cut from the 3.5 mm thick polyurethane sponge, and then a square block of 20 mm side length was cut out of the middle. Finally, the FTES was encapsulated with commercial PI membrane.

*Electrical output characterization of the FTESs*: The FTESs comprised four TENG. The open-circuit voltage and short-circuit current of the FTESs were measured using an electrometer (Keithley 6514), respectively. A LabVIEW-based software platform was developed to acquire and visualize the data in real time.

*VR demonstration*: The positioning system incorporates the ATGM332D-5N31 module, while the pulse sensor utilizes the AD8232 chip. The triboelectric signals generated by different human movements were captured by the signal acquisition module through the Arduino UNO microcontroller. The collected electric signals were transmitted to the server in real time *via* a Bluetooth module (HC-05). The received signals were then processed in Python using a trained deep model, which was also developed in Python with Keras and TensorFlow backend. The demo is developed using Three.js, an advanced framework for creating 3D interactive content. We created a cartoon character. The real-time display of corresponding actions of the cartoon character's skeleton is controlled according to the classification results produced by the neural network.

.

**Conflict of interest**

The authors declare no conflict of interest.

**Supporting Information**

Supporting Information is available from the Wiley Online Library or from the author.

**Acknowledgements**

This work was supported from the National Natural Science Foundation of China (Nos. 52072111) and Natural Science Foundation of Henan Province in China (Nos. 212300410004).

References

[1] L. The Lancet Healthy, *Lancet Healthy Longev* **2021**, 2, e180.

[2] S. Patel, H. Park, P. Bonato, L. Chan, M. Rodgers, *Journal of NeuroEngineering and Rehabilitation* **2012**, 9, 1.

[3] I. S. Chan, G. S. Ginsburg, *Annual Review of Genomics and Human Genetics* **2011**, 12, 217.

[4] T. Mishima, S. Fujioka, T. Morishita, T. Inoue, Y. Tsuboi, *Journal of Personalized Medicine* **2021**, 11, 650.

[5] R. Guo, Y. Fang, Z. Wang, A. Libanori, X. Xiao, D. Wan, X. Cui, S. Sang, W. Zhang, H. Zhang, *Advanced Functional Materials* **2022**, 32, 2204803.

[6] Z. Wang, M. Bu, K. Xiu, J. Sun, N. Hu, L. Zhao, L. Gao, F. Kong, H. Zhu, J. Song, *Nano Energy* **2022,** 104, 107978.

[7] K. R. Pyun, J. A. Rogers, S. H. Ko, *Nature Reviews Materials* **2022**, 7, 841.

[8] M. Sparkes, *New Scientist* **2021**, 251, 18.

[9] Y. Zhou, X. Xiao, G. Chen, X. Zhao, J. Chen, *Joule* **2022**, 6, 1381.

[10] S. Gannouni, A. Aledaily, K. Belwafi, H. Aboalsamh, *Scientific Reports* **2021**, 11, 7071.

[11] T. Asakawa, K. Sugiyama, T. Nozaki, T. Sameshima, S. Kobayashi, L. Wang, Z. Hong, S. Chen, C. Li, H. Namba, *Neurologia medico-chirurgica* **2019**, 59, 69.

[12] J. M. Mota, R. Baena-Pérez, I. Ruiz-Rube, M. J. P. Duarte, A. Ruiz-Castellanos, J. M. Correro-Barquín, presented at 2021 International Symposium on Computers in Education (SIIE) **2021**.

[13] D. R. Seshadri, R. T. Li, J. E. Voos, J. R. Rowbottom, C. M. Alfes, C. A. Zorman, C. K. Drummond, *NPJ Digital Medicine* **2019**, 2, 72.

[14] D. Dias, J. Paulo Silva Cunha, *Sensors* **2018**, 18, 2414.

[15] J. L. Adams, K. Dinesh, M. Xiong, C. G. Tarolli, S. Sharma, N. Sheth, A. J. Aranyosi, W. Zhu, S. Goldenthal, K. M. Biglan, E. R. Dorsey, G. Sharma, *Digit Biomark* **2017**, 1, 52.

[16]  I. H. Sarker, *SN Computer Science* **2021**, 2, 420.

[17]  A. R. Durmaz, M. Müller, B. Lei, A. Thomas, D. Britz, E. A. Holm, C. Eberl, F. Mücklich, P. Gumbsch, *Nature Communications* **2021**, 12, 6272.

[18]  R. Bagherzadeh, S. Abrishami, A. Shirali, A. R. Rajabzadeh, *Materials Today Sustainability* **2022**, 20, 100233.

[19]  A. Khan, S. Ginnaram, C.-H. Wu, H.-W. Lu, Y.-F. Pu, J. I. Wu, D. Gupta, Y.-C. Lai, H.-C. Lin, *Nano Energy* **2021**, 90, 106525.

[20]  T. Cheng, J. Shao, Z. L. Wang, *Nature Reviews Methods Primers* **2023**, 3, 39.

[21]  J. Shao, M. Willatzen, Z. L. Wang, *Journal of Applied Physics* **2020**, 128, 111101.

[22]  X. Cao, Y. Xiong, J. Sun, X. Xie, Q. Sun, Z. L. Wang, *Nano-Micro Letters* **2023**, 15, 14.

[23]  A. Pantelopoulos, N. G. Bourbakis, *IEEE Transactions on Systems, Man, and Cybernetics, Part C (Applications and Reviews)* **2009**, 40, 1.

[24]  S. An, X. Pu, S. Zhou, Y. Wu, G. Li, P. Xing, Y. Zhang, C. Hu, *ACS Nano* **2022**, 16, 9359.

[25]  W. Guo, W. T. Peter, *Journal of sound and vibration* **2013**, 332, 423.

[26]  Y. Lei, J. Lin, Z. He, M. J. Zuo, *Mechanical systems and signal processing* **2013**, 35, 108.

[27]  M. Ali, R. Prasad, *Renewable Sustainable Energy Rev.* **2019**, 104, 281.

[28]  Y. A. Shiferaw, *Physica A* **2019**, 526, 120807.

[29]  Z. Wu, N. E. Huang, *Advances in adaptive data analysis* **2009**, 1, 1.

[30]  N. E. Huang, Z. Shen, S. R. Long, M. C. Wu, H. H. Shih, Q. Zheng, N.-C. Yen, C. C. Tung, H. H. Liu, *Proceedings of the Royal Society of London. Series A: mathematical, physical and engineering sciences* **1998**, 454, 903.

[31]  L. Tarisciotti, P. Zanchetta, A. Watson, S. Bifaretti, J. C. Clare, P. W. Wheeler, *IEEE Transactions on Industrial Electronics* **2014**, 61, 6157.

[32]  K. Gao, Q. Wang, L. Xi, *Int. Arab J. Inf. Technol.* **2014**, 11, 268.

[33]  E. S. Brunette, R. C. Flemmer, C. L. Flemmer, presented at 2009 4th

International Conference on Autonomous Robots and Agents **2009**.

[34] Y. LeCun, Y. Bengio, G. Hinton, *Nature* **2015**, 521, 436.

[35] J. Ren, H. Wei, Z. Zou, T. Hou, Y. Yuan, J. Shen, X. Wang, *Power System Protection and Control* **2022**, 50, 108.

[36] J. Zhang, Y. Peng, B. Ren, T. Li, *Algorithms* **2021**, 14, 208.

[37] J. Wang, J. Li, X. Wang, T. Wang, Q. Sun, *Environment, Development and Sustainability* **2022**, 1.

[38] J. Zhang, L. Ye, Y. Lai, *Mathematics* **2023**, 11, 1985.

[39] A. Sherstinsky, Physica D 2020, 404, 132306.

[40] T. Ma, G. Xiang, Y. Shi, Y. Liu, *Geomechanics and Geophysics for Geo-Energy and Geo-Resources* **2022**, 8, 152.

[41] Z. Zhang, T. He, M. Zhu, Z. Sun, Q. Shi, J. Zhu, B. Dong, M. R. Yuce, C. Lee, *npj Flexible Electronics* **2020**, 4, 29.

[42] F. Wen, Z. Zhang, T. He, C. Lee, *Nature communications* **2021**, 12, 5378.

[43] Q. Qiu, D. A. Ramirez, S. Saleh, G. G. Fluet, H. D. Parikh, D. Kelly, S. V. Adamovich, *Journal of neuroengineering and rehabilitation* **2009**, 6, 1.

[44] T. Asakawa, K. Sugiyama, T. Nozaki, T. Sameshima, S. Kobayashi, L. Wang, Z. Hong, S. Chen, C. Li, H. Namba, *Neurologia Medico-Chirurgica* **2019**, 59, 69.

[45] J. Wang, P. Cui, J. Zhang, Y. Ge, X. Liu, N. Xuan, G. Gu, G. Cheng, Z. Du, *Nano Energy* **2021**, 89, 106320.

[46] R. Xu, F. Luo, Z. Zhu, M. Li, B. Chen, *ACS Applied Electronic Materials* **2022**, 4, 4051.

[47] R. Guo, Y. Fang, Z. Wang, A. Libanori, X. Xiao, D. Wan, X. Cui, S. Sang, W. Zhang, H. Zhang, J. Chen, *Adv. Funct. Mater*. **2022**, 32, 35.

[48] K. Song, R. Zhao, Z. L. Wang, Y. Yang, *Advanced Materials* **2019**, 31, 1902831.

[49] J. Wang, J. Jiang, C. Zhang, M. Sun, S. Han, R. Zhang, N. Liang, D. Sun, H. Liu, *Nano Energy* **2020**, 76, 105050.

[50] J. Kim, M. Jang, G. Jeong, S. Yu, J. Park, Y. Lee, S. Cho, J. Yeom, Y. Lee, A. Choe, *Nano Energy* **2021**, 89, 106409.

[51] K. Keum, J. Eom, J. H. Lee, J. S. Heo, S. K. Park, Y.-H. Kim, *Nano Energy* **2021**, 79, 105479.

[52] C. Lv, C. Tian, J. Jiang, Y. Dang, Y. Liu, X. Duan, Q. Li, X. Chen, M. Xie, *Advanced Science* **2023**, 10, 2206807.

[53] B. Ji, Q. Zhou, B. Hu, J. Zhong, J. Zhou, B. Zhou, *Advanced Materials* **2021,** 33, 2100859.

**Table 1.** Comparison of the anti-interference performance between previously proposed sensors and the proposed FTES in this work.

| Sensors | Materials | Deep learning enabled accuracy | Response time | Stability (cycle) | References |
|---|---|---|---|---|---|
| Triboelectric | PDMS/ EGaIn | No | 26 ms | 4000 | 45 |
| Triboelectric | PDMS | 95% (4 objects) | 26 ms | 3000 | 46 |
| Triboelectric | Hydrogel | 96.88% (6 objects) | 50 ms | 3000 | 5 |
| Piezoelectric | Ag/BTO/Ag | No | NR | ≈6000 | 48 |
| Piezoelectric | PVDF/PET | No | ≈170ms | NO | 49 |
| Piezoelectric | MXene/P(VDF-TrFE) | No | 53 ms | 5000 | 50 |
| Capacitive | ion-gel (IG) | No | 670 ms | 1000 | 51 |
| Capacitive | MWCNT/PDMS | No | 123 ms | 8000 | 52 |
| Capacitive | CIP/NdFeB/PDMS | No | 200 ms | 5000 | 53 |
| Triboelectric | Nylon/PET-Ito | 97.3% (8 objects) | 25 ms | 6000 | This work |

**Table 2.** The parameters of models.

| Layer (type) | Output Shape | Param |
|---|---|---|
| Conv_layer (Conv1D) | (None, 220, 64) | 2112 |
| Batch_normalization | (None, 220, 64) | 256 |
| (MaxPooling1D) | (None, 110, 64) | 0 |
| Conv_layer_1 (Conv1D) | (None, 110, 16) | 32784 |
| (BatchNormalization) | (None, 110, 16) | 64 |
| activation_1 (Activation) | (None, 110, 16) | 0 |
| Max_pooling_2 | (None, 55, 16) | 0 |
| Bilstm_layer_1 | (None, 55, 64) | 12544 |
| dropout (Dropout) | (None, 55, 64) | 0 |
| Bilstm_layer_2 | (None, 55, 64) | 24832 |
| dropout_1 (Dropout) | (None, 55, 64) | 0 |
| attention_layer | (None, 128) | 20480 |
| activation_2 | (None, 128) | 0 |
| dense (Dense) | (None, 8) | 1032 |